# ChatGPT in Classrooms: Transforming Challenges into Opportunities in Education

**Harris Bin Munawar[1], Nikolaos Misirlis[1]**

[1] HAN University of Applied Sciences, the Netherlands

**Abstract**

*In the era of exponential technology growth, one unexpected guest has claimed a seat in classrooms worldwide: Artificial Intelligence. Generative AI, such as ChatGPT, promises a revolution in education, yet it arrives with a double-edged sword. Its potential for personalized learning is offset by issues of cheating, inaccuracies, and educators struggling to incorporate it effectively into their lesson design. We're standing on the brink of this educational frontier, and it is clear that we need to navigate this terrain with a lot of care. This is a major challenge that could undermine the integrity and value of our educational process. So, how can we turn these challenges into opportunities?*

*When used inappropriately, AI tools can become the perfect tool for the 'cut-copy-paste' mentality, and quickly begin to corrode critical thinking, creativity, and deep understanding – the most important skills in our rapidly changing world. Teachers feel that they are not equipped to leverage this technology, widening the digital divide among educators and institutions.*

*Addressing these concerns calls for an in-depth research approach. We will employ empirical research, drawing on the Technology Acceptance Model, to assess the attitudes towards generative AI among educators and students. Understanding their perceptions, usage patterns, and hurdles is the first crucial step in creating an effective solution. The present study will be used as a process-manual for future researchers to apply, running their own data, based on the steps explained in here*

## 1. Introduction

Generative AI is here to stay. It promises a revolution to all fields that it is involved in. Education, as a lead component in societal changes, plays the most crucial role on understanding and incorporating such technology [1]. In an era that technology grows exponentially, generative AI tools, such as ChatGPT, are used more and more often by academics and students. As it happened before with new tools (Wikipedia, search engines, etc.), AI can be considered a double-edged knife [2]. Will it be used to improve education, or will be another tool to facilitate plagiarism? Teachers around the globe are struggling – and constantly educating themselves – with incorporating such tools into their lesson design. This represents a major challenge for the educational process. The questions remain: how ready are we as academics to use AI? How can we ensure that our, and our affiliations', integrity will not be undermined by AI? And finally, how can we turn the challenges to opportunities for the future of our science fields, our institutions and our students?

Artificial intelligence is already having a significant impact on students, offering among others, personalized tutoring, automated grading, and content creation [3]. These applications demonstrate the potential of artificial intelligence to transform learning environments. We can better understand the potential trajectory of artificial intelligence by looking at how previous technological advancements have shaped education.

On the other side, it is, however, difficult to integrate artificial intelligence into education. The most important topic to discuss is whether AI facilitates academic fraud [4]. Artificial intelligence tools can make it easier for students to engage in cut-copy-paste behavior, which undermines academic integrity, Furthermore, the precision and dependability of artificial intelligence-generated content are crucial. Misinformation and biases in data can lead to incorrect learning.

The readiness of educators to incorporate artificial intelligence into their teaching is another pressing issue. Many teachers feel ill-equipped to use these technologies effectively, which further widens the digital divide among educators and institutions [5]. The skill gap among educators affects their capacity to incorporate AI tools into their teaching methods, putting some pupils at a disadvantage.

Despite these challenges, artificial intelligence offers numerous opportunities to enhance education. The most enticing side effect is individualized instruction, where AI can tailor instruction to each student's requirements, learning paces, and learning styles. Furthermore, artificial intelligence can ease the burden of administrative duties, allowing educators to devote more time to instructing and guiding others [6].

A thorough investigation is essential to address these issues and tap into the potential of artificial intelligence. We will use the Technology Acceptance Model (TAM) to assess attitudes towards generative AI among educators and students, using empirical research. This research will involve understanding their perceptions, usage patterns, and hurdles, which will provide valuable insights into creating effective solutions.

## 2. Methodology

A thorough investigation plan will outline the selection of the sample, data collection techniques like surveys and interviews, and evaluation strategies. Understanding the factors that influence acceptance and use of AI in education will allow us to develop strategies to maximize its benefits while mitigating its risks [7].

Before listing the various questions related to TAM, it is important to mention that the following survey is meant to be as generic as possible [8]. Educators and researchers who will run the survey can (and must) adapt it to their needs, institutions, cultural environment, political situation, and financial status of the country they teach. Furthermore, a future research may include items related to Theory of Planned behavior. This way the paths formulated after the statistical analysis may lead to more complex but also more useful insights [9].

### *2.1 Survey*
***Section 1:***
Demographic Information
Institution Type
Field of Study/Teaching
Years of Experience (for educators)

**Section 2 (Likert items): Perceived Usefulness (PU)**
1. Using generative AI tools (e.g., ChatGPT) enhances my learning/teaching experience.
2. Generative AI makes it easier to complete academic tasks.
3. Generative AI improves the quality of my learning/teaching.
4. Generative AI allows for more efficient time management in learning/teaching.

**Section 3 (Likert items): Perceived Ease of Use (PEOU)**
1. Learning to use generative AI tools is easy for me.
2. Interacting with generative AI tools does not require a lot of mental effort.
3. I find generative AI tools to be flexible to interact with.
4. It is easy for me to become skillful at using generative AI tools.

**Section 4 (Likert items): Attitude Towards Using (ATU)**
1. Using generative AI tools is a good idea.
2. I like the idea of using generative AI tools in my learning/teaching.
3. Using generative AI tools is beneficial for my academic performance.

**Section 5 (Likert items): Behavioral Intention to Use (BI)**
1. I intend to use generative AI tools in my future learning/teaching activities.
2. I will recommend others to use generative AI tools.
3. I plan to increase my use of generative AI tools in the future.

**Section 6: Usage Patterns**
1. How frequently do you use generative AI tools (e.g., ChatGPT)?
2. For what purposes do you use generative AI tools?

**Section 7: Challenges and Concerns**
1. What challenges do you face when using generative AI tools?
2. How concerned are you about the potential for cheating and plagiarism with the use of generative AI?
3. How concerned are you about the bias and inaccuracies in AI-generated content?

### 3. Structural Equation Modelling (SEM) Explanation

Structural Equation Modeling (SEM) is a multivariate statistical analysis method used to analyze structural relationships. This technique is a combination of factor analysis and multiple regression analysis and is used to analyze the structural relationship between measured variables and latent constructs.

Steps to Apply SEM for TAM Data

1. Model Specification:
   - Define the theoretical model based on TAM. This includes the relationships between latent variables such as Perceived Usefulness (PU), Perceived Ease of Use (PEOU), Attitude Toward Using (ATU), and Behavioral Intention to Use (BI).
   - Hypothesized Relationships (Just an example. Different paths may occur) :
     - PEOU → PU
     - PU → ATU
     - PEOU → ATU
     - ATU → BI
     - PU → BI

2. Model Identification:
   - Ensure that the model is identified, meaning there are enough degrees of freedom to estimate the model. This involves ensuring that the number of observations is sufficient and that the model is not over-parameterized. Furthermore, researchers need to be sure about the number of the responses obtained.

3. Data Collection:
   - Use the TAM survey to collect data from a representative sample of educators and students.
   - Ensure data quality by checking for missing values and outliers.

4. Measurement Model:
   - Perform Confirmatory Factor Analysis (CFA) to validate the measurement model. This step involves checking whether the observed variables accurately measure the latent constructs (PU, PEOU, ATU, BI).
   - Model Fit Indices: Assess the fit of the measurement model using fit indices such as:
     - Chi-Square ($\chi^2$) Test
     - Root Mean Square Error of Approximation (RMSEA)
     - Comparative Fit Index (CFI)
     - Standardized Root Mean Square Residual (SRMR)

5. Structural Model:
   - Specify the structural model to test the hypothesized relationships between the latent constructs.
   - Estimate the model using software such as AMOS, LISREL, or Mplus.
   - Model Fit Assessment: Evaluate the overall fit of the structural model using the same fit indices as in the measurement model.

6. Path Analysis:
   - Analyze the path coefficients to determine the strength and significance of the relationships between latent constructs.
   - Direct Effects: Examine the direct effects of PEOU on PU, PU on ATU, PEOU on ATU, and ATU on BI.
   - Indirect Effects: Calculate indirect effects (e.g., PEOU → PU → ATU) if applicable.

7. Model Modification:
   - If necessary, modify the model based on fit indices and theoretical justification. This may involve adding or removing paths, correlating error terms, or re-specifying relationships.
   - Re-estimate the modified model and reassess model fit.

8. Reporting Results:
   - Present the results of the SEM analysis, including:
     - Descriptive statistics and correlations among observed variables.
     - Results of the CFA, including factor loadings and fit indices.
     - Path coefficients, standard errors, and significance levels for the structural model.

- Interpretation of the results in the context of TAM and generative AI in education.

Table 1 contains the recommended values for the indices to be measured [10].

*Table 1: Recommended values of SEM for the fit indices.*

| Fit indices | Recommended values |
|---|---|
| $x^2$/ d.f. | ≤ 3.00 |
| NNFI | ≥ 0.90 |
| RMSEA | ≤ 0.09 |
| AGFI | ≥ 0.80 |
| GFI | ≥ 0.90 |
| CFI | ≥ 0.90 |
| RMR | ≤ 0.05 |

### *3.1 Proposed structural model*

To better facilitate the future development of studies, we propose an initial structural model to use (Figure 1). The proposed model can be adapted as, external variables could be added before the PU and PEOU components. Those external variables are not standardized, and they depend on the nature of the research. If the researchers want to elaborate on the correlations between psychological factors and the TAM, then a BIG Five model could be used [11, 12]. On the other hand, different cultural factors may be analysed, or the use of other technologies together with AI [13, 14].

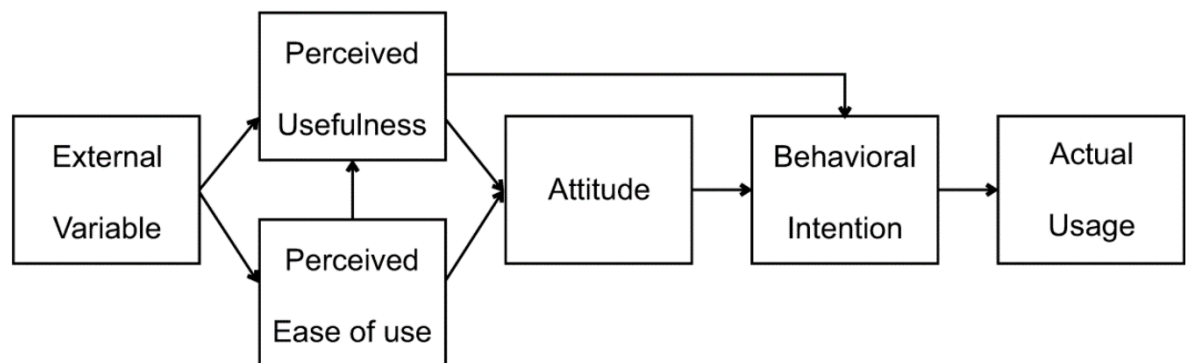

*Figure 1:Proposed structural model*

### 4. Conclusions

Several key areas need to be addressed in order for educational institutions to successfully integrate artificial intelligence into their curriculum. Education requires specific training and support to use artificial intelligence tools efficiently for both teachers and students. Using AI, educators and researchers can enhance their teaching techniques and close the knowledge gap. But *only* if they are ready to accept the upcoming change.

Future research should explore the evolving landscape of artificial intelligence in education, and most important the level of acceptance among academics. Changes in acceptance and usage patterns can be tracked over time by using our proposed process. In order to gain a more comprehensive understanding of the global impact of generative AI, it would be beneficial to expand the research to include diverse educational settings, cultural backgrounds, different economies, and large sample sizes.